\documentclass[12pt]{iopart}
\usepackage{braket}
\usepackage{graphicx}
\graphicspath{{figures/}}

\pdfoutput=1

\usepackage{ulem}

\usepackage{color}
\usepackage{subfigure}
\usepackage{gnuplottex}
\usepackage[english]{babel}

\begin{document}

\title[Generation of NOON states]{Non-adiabatic generation of NOON states in a Tonks--Girardeau gas\footnote{Dedicated to the memory of Marvin D.~Girardeau.}}

\author{J.~Schloss, A.~Benseny, J.~Gillet, J.~Swain, and Th.~Busch}

\address{Quantum Systems Unit, Okinawa Institute of Science and Technology, 904-0495 Okinawa, Japan}

\ead{james.schloss@oist.jp}

\begin{abstract}
With adiabatic techniques, it is possible to create quantum superposition states with high fidelity while exercising limited control over the parameters of a system. However, because these techniques are slow compared to other timescales in the system, they are usually not suitable for creating highly unstable states or performing time-critical processes. 
Both of these situations arise in quantum information processing, where entangled states may only be isolated from the environment for a short time and where quantum computers require high-fidelity operations to be performed quickly.
Recently it has been shown that techniques like optimal control and shortcuts to adiabaticity may be used to prepare quantum states non-adiabatically with high fidelity.
Here we present two examples of how these techniques can be used to create maximally entangled many-body NOON states in one-dimensional Tonks--Girardeau gases. 
\end{abstract}

\pacs{03.67.-a, 03.67.Bg, 42.50.Dv}

\maketitle


\section{Introduction}

Macroscopic superposition states, such as the maximally entangled $\ket{N,0} + \ket{0,N}$ (NOON) state, are of great interest for fundamental studies of quantum mechanics and for applications in quantum information and quantum metrology. A NOON state is composed of two modes where all particles in the system can be found in either one mode or the other. Until now, experimental NOON state generation has been limited to photonic states generated by mixing classical states with down-converted photon pairs~\cite{EXP1}, and with such techniques, it is possible to create NOON states with around 5 photons ~\cite{EXP5}.
A theoretical proposal for an experimentally realistic setup for creating NOON states for a large number of ultracold atoms was recently presented by Hallwood \etal~\cite{RING}, who considered a gas of strongly interacting bosons in a one-dimensional ring. In this proposed system, states with different angular momentum may become coupled by breaking the rotational symmetry, and the authors have shown how to accelerate the atoms into a superposition state of rotating and non-rotating components. Furthermore, since the atoms are considered to be in the strongly correlated (Tonks--Girardeau~\cite{Girardeau}) regime, this process results in a macroscopically-entangled state.

In order to successfully generate NOON states on a ring of strongly correlated ultracold atoms, it is crucial to rotationally accelerate the system slowly, as otherwise unwanted excitations may drive the system out of the desired state. This is especially important close to the avoided crossings where the NOON state lives and where states with different angular momentum quantum numbers are coupled. For larger particles numbers and finite width coupling barriers, the energy gaps at these positions become exponentially small~\cite{BARRIERS}, and therefore slower and slower driving is necessary. However, slow processes are not particularly suitable for applications in quantum information, where quantum algorithms must be performed quickly, or for creating states that are highly unstable and techniques which can speed up the creation process  while  maintaining high fidelities are of large interest.

Here we present two examples of such techniques that can accelerate the creation of the NOON states.
The first is the Chopped RAndom Basis (CRAB) optimal control technique~\cite{CRAB}, where we numerically optimise the angular acceleration and the height of a barrier.
The second technique combines two well-known shortcuts to adiabaticity (STA) protocols \cite{STA} which we adapt to the ring geometry.  
In both cases we show that it is possible to drive the system suggested by Hallwood \etal~\cite{RING} into a NOON state on timescales much faster than required by adiabaticity.

The paper is organised as follows: In Section 2 we begin briefly review the ring system of strongly correlated ultracold atoms proposed by Hallwood \etal~\cite{RING}. This is followed in Section 3 by a detailed description of how to use the CRAB algorithm to create NOON states non-adiabatically with optimal control techniques.  In Section 4, we show how Shortcuts to Adiabaticity (STA) can be used to create NOON states with high fidelity in a similar system. We finish with the conclusions in Section 5.

\section{Creating a NOON state on a ring}

\begin{figure}
\begin{center}
\includegraphics[width=0.4\textwidth]{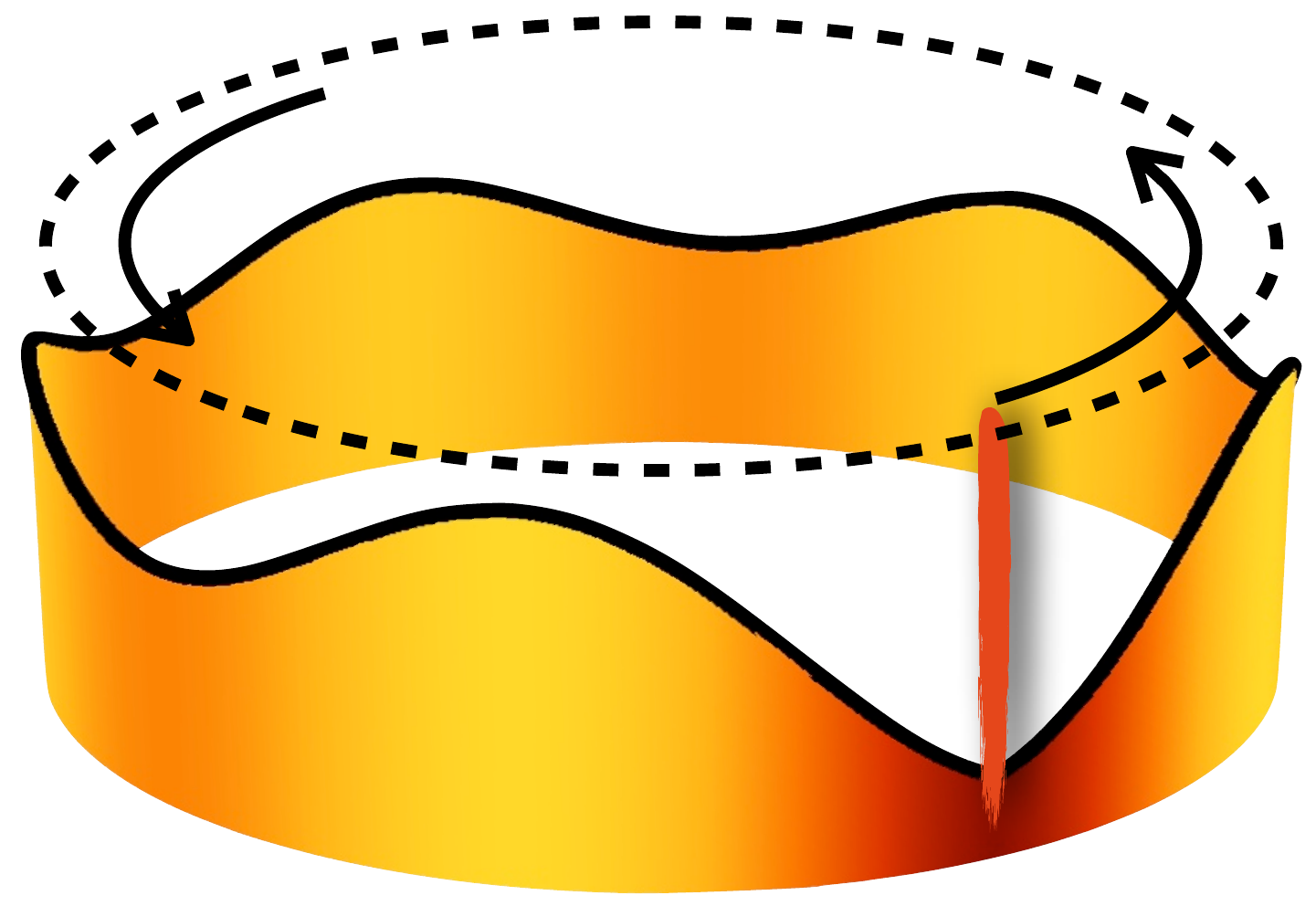} 
\caption{
\label{fig:Scheme}
 Schematic of the system described by Hallwood \etal~\cite{RING}. Shown is the density profile for 5 atoms in the TG regime which are stirred by a highly localised potential (indicated by the vertical line).} 
\end{center}
\end{figure}

Let us begin by briefly summarising the protocol suggested by Hallwood \etal~\cite{RING, ENGINEERING} for creating a NOON state in a gas of strongly correlated bosons.
For this we consider a gas of $N$ interacting bosons of mass $m$ on a one-dimension ring with circumference $L$.
The system includes a potential barrier modelled by a Dirac $\delta$ function, rotating with angular frequency $\Omega$;
a schematic can be seen in Figure~\ref{fig:Scheme}.
In the rotating frame,  the Hamiltonian is given by \cite{RING}
\begin{equation}
H^{(N)} = \sum_{i=1} ^{N} \left[{\frac{1}{2}\bigg(-i\frac{\partial}{\partial x_i}-\Omega}\bigg)^2 + b\delta(x_i) +g \sum_{i<j} ^{N} \delta (x_i - x_j )\right],
\end{equation}
where, $b$ is the height of the barrier (in units of $\hbar^2/mL^2$), $x_i \in \left[-1/2,1/2\right]$ is the position of the $i$--th particle (in units of $L$) and $g$ (in units of $\hbar^2/mL^2$) is the effective interaction strength between the atoms. 

In the strongly-correlated Tonks--Girardeau (TG) limit ($g \rightarrow \infty$),
the Hamiltonian $H^{(N)}$ can be solved by using the Bose--Fermi mapping theorem \cite{girardeau2001,girardeau2000}.
From this, it is possible to calculate the stationary states and the time evolution of the system by mapping the bosonic atoms to non-interacting fermions and symmetrising the resulting many-particle wavefunction afterwards. The time evolution of the entire system therefore only requires evolving single-particle states, governed by the  laboratory-frame Hamiltonian 
\begin{equation}
H = -\frac{1}{2} \frac{\partial^2}{\partial x} + b\delta \left[ x-x_0(t) \right], 
\end{equation}
where $x_0$ is the position of the barrier.

\begin{figure}
 \centering
 \subfigure{
 \centering
 \includegraphics[width = 0.4\linewidth]{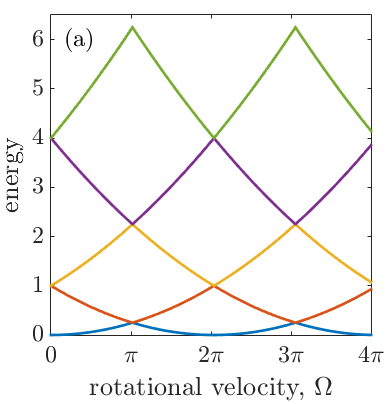}} 
 \subfigure{
 \centering
 \includegraphics[width = 0.4\linewidth]{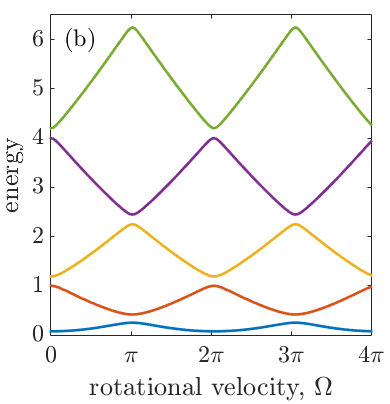}}
 \caption{
 \label{fig:avoid}
 Single particle energy spectrum as a function of $\Omega$ for a barrier height of (a) $b=0$ and (b) $b=2$. The presence of the barrier leads to the appearance of avoided crossings at intervals of $\pi$, which grow in size as a function of barrier height. The degenerate eigenstates for $b=0$ when $\Omega$ is multiple of $2\pi$ correspond to clockwise and counterclockwise rotating plane waves.}
\end{figure}

The energy spectrum of this system is shown in Figure~\ref{fig:avoid} as a function of the rotational velocity $\Omega \equiv \dot x_0 /L$ (in units of $\hbar/mL$) of the system. In the absence of barrier, $b = 0$ (Figure~\ref{fig:avoid}(a)), the eigenstates of $H$ are plane waves with quantised angular momentum in units of integer multiples of $2 \pi$ and manifolds of fixed angular momentum are uncoupled due to the existence of rotational symmetry.
However, when $b > 0$ (Figure~\ref{fig:avoid}(b)) this symmetry is broken and transitions between different manifolds become possible \cite{CROSSING, SUPERPOSITION}, resulting in the avoided crossings shown in the energy spectrum.

By adiabatically accelerating the barrier from $\Omega = 0$ to $\pi$, a particle initially in an eigenstate of $H$ will enter a superposition of two angular momentum eigenstates.
However, any non-adiabatic behaviour will lead to transitions to higher or lower lying states in the vicinity of the gaps which would degrade the fidelity of the superposition state.
Since in the TG limit, the strongly correlated many-particle wavefunction can be directly calculated from the single particles ones, we may create a macroscopic NOON state between successive values of angular momentum with this process~\cite{RING}. 
The condition for adiabaticity of this system must therefore be chosen with respect to the smallest gap size, which in general decreases exponentially for higher energies~\cite{GENERATION}.
For a delta barrier, however, the gap size can be shown to stay constant to first order~\cite{BARRIERS}.

To avoid the restrictions set by adiabatic evolution, we consider two strategies in the following. The first is an optimal control technique, which determines an optimal form (``pulse'') of the non-adiabatic rotational velocity $\Omega(t)$ through brute-force computational methods, which will generate the desired final state with high fidelity. For this we have implemented the CRAB optimal control technique~\cite{CRAB}, which starts by initially assuming a constant acceleration of the angular barrier velocity from $\Omega=0$ to $\pi$ and iterates on this by including procedurally generated sinusoidal variations.
After each iteration, the fidelity is calculated and the Nelder--Mead method \cite{SIMPLEX} is used to find the pulse that gives a final state closest to the desired one. This ultimately leads to a form of the rotational velocity that maximises the fidelity after reaching the NOON state in a preset amount of time.
In a similar way, we also find optimal pulses for the barrier height, and for a combination of both the rotational velocity and the barrier height.

In contrast, the second strategy we consider combines two known results from the area of STAs~\cite{STA}. 
For this we break the process into two parts: a first one which breaks the rotational symmetry (raising of a potential), and a second which accelerates the atoms.
For both of these tasks shortcuts are known for atoms trapped in a harmonic oscillator potential~\cite{STA}, and we assume that such a potential can be created along the perimeter.
In order to maximise the entanglement of the final state, we will also consider restoring the rotational symmetry by lowering the potential at the end.
It is worth mentioning that a fast quasi-adiabatic (FAQUAD) shortcut for creating a TG gas superposition state as described above has been recently developed~\cite{Gillet}.

To quantify the success of our protocols, we use the fidelity $F = \left| \langle \psi | \phi \rangle \right|^2$, where 
 $\ket{\psi}$ is the achieved state and $\ket\phi$ is the target state.
When $F$ is close to one, it is convenient to also define the {\it infidelity} as $1-F$.
The fidelity between two many-particle TG states, 
$\ket{\Psi}$ and $\ket{\Phi}$, can be calculated by using the mode by mode projections 
\begin{eqnarray}
\braket{\Psi | \Phi} &=& \frac1{N!} \sum_{\eta, \mu} \epsilon_\eta \epsilon_\mu \braket{\psi_{\eta_1}(x_1) | \phi_{\mu_1}(x_1) } \cdots \braket{ \psi_{\eta_N}(x_N) | \phi_{\mu_N}(x_N) } \nonumber \\
\label{eq:fid}
&=& \det \left(
\begin{array}{ccc}
\braket{\psi_1 | \phi_1 } & \dots & \braket{\psi_1 | \phi_N } \\
 \vdots & \ddots & \vdots \\
\braket{\psi_N | \phi_1 } & \dots & \braket{\psi_N | \phi_N } 
\end{array}
\right),
\end{eqnarray}
which follows directly from the form of the TG state
\begin{equation}
\Psi(x_1,x_2,\ldots, x_N)= \frac1{\sqrt{N!}} \prod_{i<j}\textnormal{sign}(x_i-x_j) \sum_{\eta \in P} \epsilon_\eta \psi_{\eta_1}(x_1)\cdots\psi_{\eta_N}(x_N).
\label{eq:TG}
\end{equation}
Here $P$ represents the set of all permutations of $\{0,1 \dots N-1\}$, $\epsilon_\eta$ represents the antisymmetric tensor of the permutation $\eta$, and $\psi_i$ represent the orbitals. These definitions will be used to measure how close the final state of our finite time algorithms comes to the perfect NOON state. 

\section{Optimal Control}

Let us now discuss an optimal control approach for generating NOON states using the system proposed by Hallwood \etal~\cite{RING}. For this we implement the Chopped RAndom Basis (CRAB) optimal control algorithm \cite{CRAB, OC} for systems of up to 5 particles. It should be mentioned that unlike some other studies on the optimising 1D bosonic ring systems \cite{OPTIMAL, OPT_SCALE}, here we focus on fast, non-adiabatic creation of NOON states. The CRAB technique works by modifying a control parameter of a given system, $\Gamma$, with a multiplicative term as
\begin{equation}
 \Gamma^{\rm CRAB}(t) = \Gamma^0(t)\gamma(t) ,
\end{equation}
where $\Gamma^0(t)$ is an initial guess, and the function $\gamma(t)$
is written as a sum of sinusoidal functions
\begin{equation}
\gamma(t) = 1 + \frac{\sum^{J}_{j = 0}A_{j}\sin(\nu_{j}t)+B_{j}\cos(\nu_{j}t)}{\lambda(t)} .
\end{equation}
To ensure that $\Gamma^{\rm CRAB}(t)$ and $\Gamma^0(t)$ coincide at the initial and final times, $\lambda(t)$ is defined such that
$\lim_{t \rightarrow 0} \lambda(t) = \lim_{t \rightarrow T} \lambda(t) = \infty$.
In our implementation we chose
\begin{equation}
\lambda(t) = \frac{T^2}{4t(t-T)} .
\end{equation}

The optimisation process then reduces to finding the optimal values for $\{ A_{j}, B_{j}, \nu_{j} \}$, which can be achieved by initially assigning random values and then numerically maximising the fidelity by using an algorithm such as the Nelder--Mead method \cite{SIMPLEX}.
While it is clear that this process will lead to more accurate outcomes for larger $J$, the fact that the maximisation has to be carried over a larger number of degrees of freedom also increases the computational complexity.
In our case, $\Gamma (t)$ can be chosen to be the rotational angular frequency or the barrier height, which means that we may optimise over the rotation frequency while keeping the barrier height constant, optimise over the barrier height while keeping the rotation acceleration constant, or optimise over both of them simultaneously. These three possibilities will be expanded upon in the next sections. 
 
 \subsection{Optimising over the rotational velocity}
 
 \begin{figure}[t]
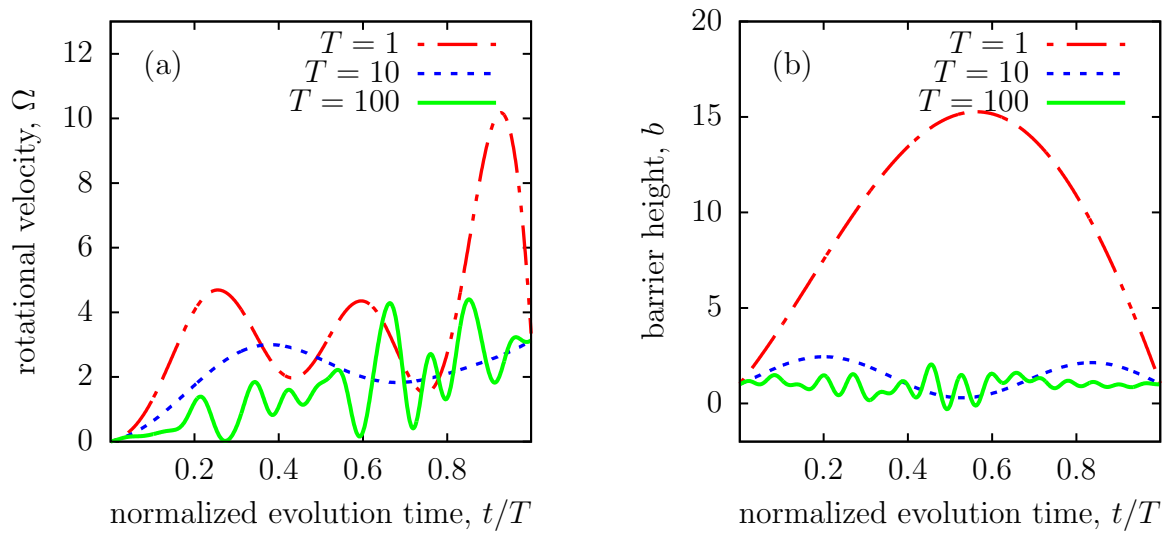

 \subfigure{
 \centering
 \input{figR0.tex}
 }
 \subfigure{
 \centering
 \input{figB0.tex}
 }
 \caption{ (a) Optimal rotational velocity pulses for $T = 1$, 10, and 100 for fixed barrier height $b=1$. (b) Optimised barrier height for a linearly increasing rotational velocity, $\Omega = \pi t / T$ for $T=1$, 10, and 100. }
 \label{fig:pulses}
\end{figure}

To find the optimal pulse for the rotational velocity in the presence of a fixed barrier height, we modify a linear guess pulse, starting at $\Omega = 0$ and reaching $\pi$ in a preset total time $T$.
The results for $J=15$ and for $T=1$, 10, and 100 are shown in Figure~\ref{fig:pulses}(a).
For longer evolution times, a linear pulse is a reasonable method to adiabatically generate the macroscopic superposition state with high fidelity, so the deviations stemming from the optimal control process for $T=100$ can be seen to be weak in magnitude. 
Shorter evolution times, however, require pulse shapes that strongly influence the system and therefore differ dramatically from the initial linear guess.
From the infidelities for the linear guess pulse and the optimised pulse, shown in Figure~\ref{fig:lfid}, one can see an improvement of several orders of magnitude on all timescales.

\subsection{Optimising over the barrier height}

\begin{figure}
 \centering
	\input{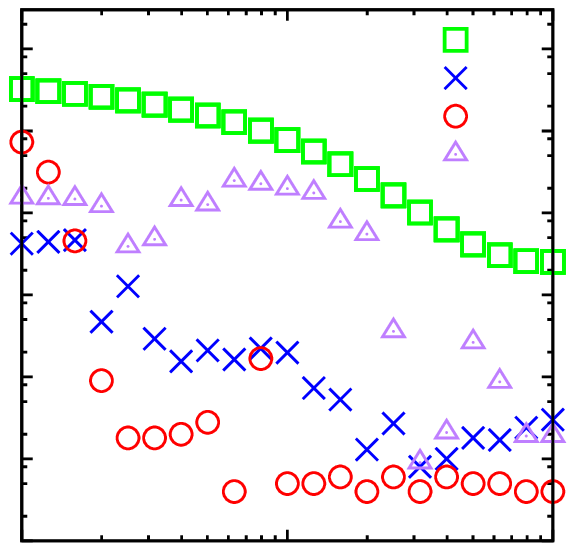}
\caption{ Infidelities as a function of the overall process time for optimised control of the rotational acceleration, the barrier height, or both. Here, ``linear'' refers to an unoptimised linear acceleration from $\Omega = 0$ to $\pi$ while keeping the barrier height fixed at $b=1$. 
}
\label{fig:lfid}
\end{figure}

To optimise over the barrier height, we choose a guess pulse which is constant at $b=1$ while the rotational velocity of the barrier is set to increase linearly from $\Omega = 0$ to $\pi$ over a total time $T$.
The optimised pulses for the barrier heights for $T = 1$, 10, and 100 for $J=15$ are shown in Figure~\ref{fig:pulses}(b), and,
similar to the case of varying $\Omega$, shorter evolution times require larger deviations from the initial guess.
As in the previous case, these pulses also lead to significant improvements in the final fidelity, shown in Figure~\ref{fig:lfid}.

\subsection{Optimising over rotational velocity and barrier height}

\begin{figure}
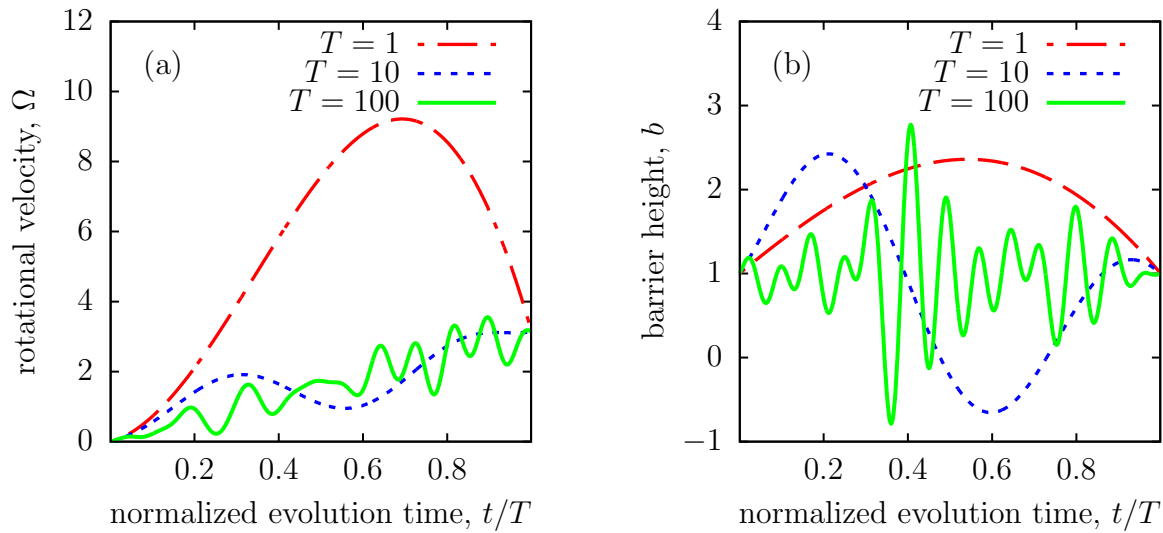
 
 \subfigure{
 \centering
 \input{figR1.tex}
 }
 \subfigure{
 \centering
 \input{figB1.tex}
 }
 \caption{Optimal pulses for $T = 1$, 10, and 100 of the (a) rotational velocity and (b) barrier height found when optimising over both simultaneously. }
 \label{fig:pulses_pair}
\end{figure}

With the CRAB algorithm, it is possible to optimise over multiple parameters at the same time. In Figure~\ref{fig:pulses_pair}, we show the optimal pulses for simultaneously changing the rotational velocity and barrier height.
Compared to the previous cases, where only one parameter was optimised, one can see that for longer evolution times the resulting pulse shapes are similar.
For shorter times, however, they differ significantly (compare the red lines in Figs.~\ref{fig:pulses} and \ref{fig:pulses_pair}); however, it would seem that the final fidelity is not drastically different from optimizing over the rotation of the barrier, alone. From this, we may conclude that it is sufficient to optimize only over the rotation of the barrier to optimize the fidelity of our system. 

\subsection{Tonks--Girardeau gas acceleration}

\begin{figure}[t]
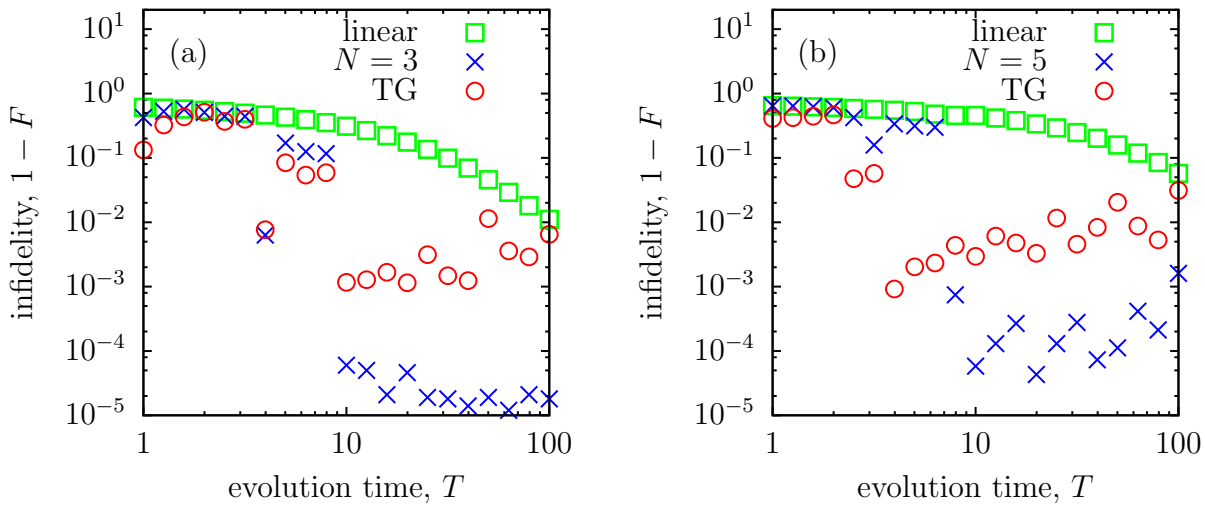

 \subfigure{
 \centering
 \input{figTG3.tex}
 }
 \subfigure{
 \centering
 \input{figTG5.tex}
 }
 \caption{ Infidelities for the TG gas evolution with the CRAB optimal control technique with $N=3$ (a) and $N=5$ (b) particles. In these simulations, we optimised over the fidelity of the particle with highest energy. These simulations show a clear range where the CRAB algorithm is effective for generating NOON states with multiple particles.}
 \label{fig:TGOC}
\end{figure} 

Due to the Bose--Fermi mapping theorem, the evolution of an $N$-particle TG gas can be calculated by evolving a Fermi sea of $N$ spin-polarised independent fermions. Transitions into empty levels above the Fermi edge will affect the global fidelity, and therefore the particle at the Fermi edge is the most crucial one for this parameter~\cite{Gillet}. Thus, we have used the CRAB algorithm to determine the optimal rotational velocity pulse to generate a NOON state in the TG gas by optimising the fidelity of the particle at the Fermi edge.
In Figure \ref{fig:TGOC} we show the fidelity for both the $N$-th particle and the entire TG gas (for $N=3$ and 5) and one can see that
 that CRAB used this way gives highly effective pulses for all situations. 
Note that for very short and long evolution times, the fidelity increase from the CRAB algorithm for a TG gas is less apparent with more than one atom. 

\section{Shortcuts to adiabaticity}

\begin{figure}
\centering
\includegraphics[width=0.65\linewidth]{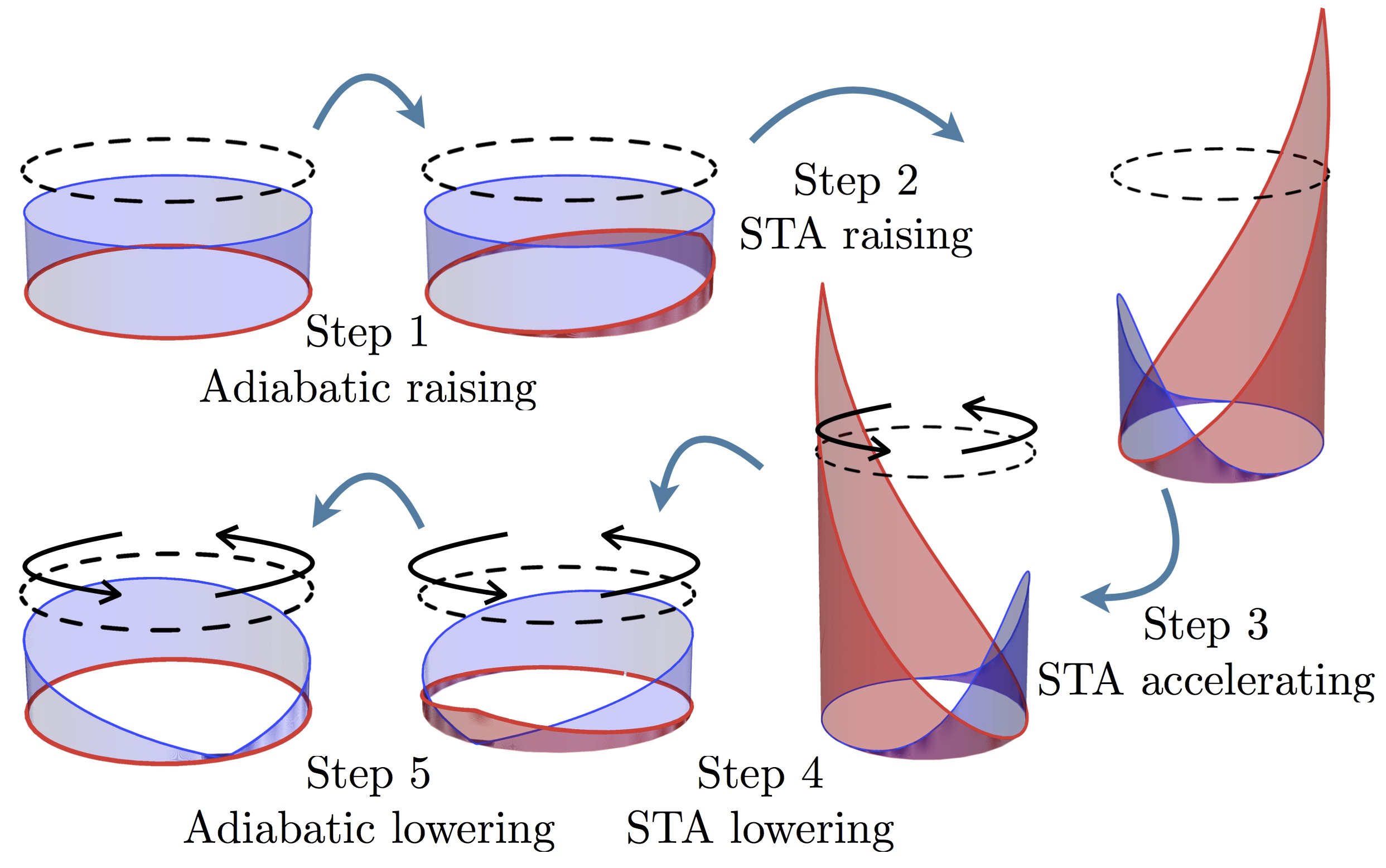} 
\caption{Scheme for the acceleration of a single atom using STA
 In this example, the ground state of free space gets tightened, accelerated and released at the angular velocity of $\Omega=\pi$ into the state
 $\left( \exp(i 2\pi x) +1 \right) /\sqrt 2$}
\label{fig:STA-scheme}
\end{figure}

In the following we will investigate the possibility of creating a NOON state by using STA techniques~\cite{STA}. For this we manipulate the eigenstates of the ring system by introducing time-dependent external potentials that we ultimately remove.
This leaves the system in a NOON state without the need for a potential barrier, which is different from the set-up proposed by Hallwood \etal~\cite{RING}.

The protocol we suggest consists of five steps, (1) adiabatically raising a weak harmonic potential wrapped around the ring, (2) quickly tightening this potential via a shortcut to localise the particle, (3) accelerating the particle by moving the centre of the harmonic potential via another shortcut, (4) lowering the potential via the reverse process of step (2) to delocalise it again, and (5) adiabatically removing the harmonic potential.
A schematic of this process is shown in Figure~\ref{fig:STA-scheme}.
In the next section, we will briefly review the general framework of the STA formalism~\cite{STA} and detail the differences of our protocol with respect to existing ones.

\subsection{Lewis--Riesenfeld invariants}

STA methods based on the Lewis--Riesenfeld invariant inverse-engineering approach~\cite{LEWIS} make use of the existence of invariants.
A one-dimensional Hamiltonian has an invariant quadratic in momentum $p$ if and only if it can be expressed in the following manner
\begin{equation}
H = \frac{p^2}{2m} -F(t)q+\frac{m}{2}\omega^2(t)q^2+\frac{1}{\rho(t)^2}U\left(\frac{q-q_c(t)}{\rho(t)}\right), 
\label{eq:HSTA}
\end{equation}
where $U$ is an arbitrary function and $q$ is the position operator. 
The variables $\rho$, $q_c$, $\omega$, and $F$ are arbitrary functions of time satisfying the auxiliary equations
\begin{eqnarray}
\ddot{\rho}+\omega^2(t)\rho =\frac{\omega_0^2}{\rho^3}, \label{eq:rho} \\
\ddot{q}_c+\omega^2(t)q_c = F(t)/m, \label{eq:qc}
\end{eqnarray}
where $\omega_0$ is a constant.
The physical interpretation of the constant and functions depends on the underlying system.
Additional constraints have to be considered to insure that the Hamiltionian and its invariants commute at initial and final times $t_0$ and $t_f$, which in our case results in $H(t_0)=H(t_f)=p^2/2m$.

\subsection{Shortcut for raising/lowering the potential}

One of the two shortcuts being used in the protocol above involves raising or lowering a harmonic potential. For this we only need a stationary harmonic potential and can set $F$, $q_c$ and $U$ from Eq.~(\ref{eq:HSTA}) to zero, which leads to 
\begin{equation}
 H= -\frac{1}{2} \frac{\partial^2}{\partial x}+ \frac 1 2 \omega^2(t) q^2,
\end{equation}
complemented by the single auxiliary equation (\ref{eq:rho}).
To change the frequency from $\omega(t_0)=\omega_0$ to $\omega(t_f)=\omega_f$ while keeping the commutation relations and $\omega(t)$ continuous, we impose the conditions
\begin{equation}
 \begin{array}{lcl}
\rho(t_0)=1, && \rho(t_f)=\gamma=\sqrt{\omega_0 / \omega_f},\\
\dot \rho(t_0)=0, && \dot \rho(t_f) =0, \\
\ddot \rho(t_0)=0, && \ddot \rho(t_f)=0.
\end{array} \label{eq:squeeze}
\end{equation}
This means that as long as the conditions (\ref{eq:rho}) and (\ref{eq:squeeze}) are obeyed, we are free to choose any specific form of $\rho$. A good choice is a simple polynomial of the form
\begin{equation}
 \rho (s) = 6 \left(\gamma -1\right) s^5 -15 \left(\gamma-1\right) s^4 +10 \left(\gamma-1\right)s^3 + 1, \label{eq:rho_pol}
\end{equation}
where $s=(t-t_0)/(t_f-t_0)$, which, when inserted into Eq.~(\ref{eq:rho}), allows us to numerically find a solution for $\omega(t)$. This solution leads to the necessary squeezing or expansion of the particle wavefunction with perfect fidelity in an arbitrarily short time $t_f-t_0$, although in practice a shorter squeezing time involves a faster variation of $\omega$, which may be limited by technical capabilities. Note that even though this shortcut was not specifically built for periodic boundary systems it can also be used in a ring, as the symmetry of the potential is never broken during the time evolution.

It is also important to note that the initial frequency $\omega_0$ can be chosen arbitrarily small as long as it is non-zero.
However, the solution of Eq.~(\ref{eq:rho}) with very small values of $\omega_0$ can yield purely imaginary values of $\omega(t)$, which corresponds to inverted (repulsive) potentials. While changing potentials between attractive and repulsive is experimentally possible, such a procedure is often associated with very fast and large changes, which may not be easy to realise experimentally. Furthermore, our desired initial and final potential do require an exact $\omega=0$, which forces us to include the first step to our protocol of raising $\omega$ from 0 to a suitable $\omega_0$ adiabatically slowly before being able to use the shortcut protocol. In a similar manner, the last step of lowering the potential to $\omega=0$ after having accelerated the particle has to be done adiabatically.

\subsection{Shortcut for the acceleration}

Once the potential is raised, the next step is to accelerate the particle.
A shortcut exists for this process in a harmonic trap, with which we may keep the trapping frequency constant at $\omega_0$, and only requires a change in the the position of the potential.
We can therefore set $U=0$ and $F=\omega_0^2 x_0(t)$ which leaves
\begin{equation}
 H= -\frac{1}{2} \frac{\partial^2}{\partial x}+ \frac 1 2 \omega^2_0 (q-x_0(t))^2, 
\end{equation}
and condition (\ref{eq:qc}) becomes the only relevant auxiliary equation
\begin{equation}
 \ddot{q}_c+\omega^2_0 (q_c-x_0)=0.
\end{equation}
We once again impose conditions on $q_c$ such that all boundary conditions are satisfied
\begin{equation}
 \begin{array}{lcl}
q_c(t_0)=x_0(t_0), && q_c(t_f)=d,\\
\dot q_c(t_0)=0, && \dot q_c(t_f) =\Omega_f, \\
\ddot q_c(t_0)=0, && \ddot q_c(t_f)=0,
\end{array}
\end{equation}
where $d$ is the final position of the potential minimum and $\Omega_f$ is its final velocity. For transport schemes, $d$ is the important parameter and $\Omega_f$ is set to 0, but in our case the opposite occurs as we want the particle to accumulate kinetic energy before being released from the potential (that keeps revolving at constant speed $\Omega_f$ after $t_f$) and the final position $d$ plays no significant role in the evolution of our states.

Again, the exact form of $q_c$ can be chosen arbitrarily and we pick the following polynomial
\begin{equation}
 q_c(s)= (6 d -3 \Omega_f )s^5 - (15 d-7 \Omega_f )s^4+(10d-4 \Omega_f) s^3 + x_0(t_0),
\end{equation}
where, as above, $s$ is the normalised time.
The value of $\Omega_f$ can be chosen as a multiple of $2\pi$ to create a plane wave after release, or as an odd multiple of $\pi$ to prepare superpositions between states of different angular momentum.

Note that this transport scheme is usually applied to an open, infinite space, whereas our system has periodic boundary conditions.
Since translational symmetry is broken, the shortcut is no longer guaranteed to work perfectly, as the potential has a finite height and therefore higher-lying states are no longer trapped.
Therefore, unlike the potential raising shortcut, this accelerating shortcut is only approximate and works best when $\omega$ is large (the particle is highly localised) and the rotational velocity, $\dot q_c$, is not too high (the harmonic well is not moving too fast).

\subsection{Harmonic and sinusoidal potentials}

Both shortcuts described above are based on the presence of a harmonic potential of the form
\begin{equation}
 V_{H}(x,t)=\frac 1 2 \omega^2(t) \left( x-x_0(t)\right)^2, 
\end{equation}
where  $\omega$ is the frequency of the trap (in units of $\hbar/mL^2$) and $x_0$ the position of its minimum.
Note that we require the potential to be symmetric around $x_0$ so that the potential is continuous at $x=\pm 1/2$,  and therefore the real form of $(x-x_0)$ must be $(x-x_0+1/2)(\mathrm{mod~} 1)-1/2$.
The potential $V_H$ is then continuous everywhere on the ring, but its derivative is discontinuous at $x=x_0+1/2$ (the position diametrically opposite to $x_0$). From a theoretical perspective, $V_H$ is ideal because of its simplicity and its numerous known properties, particularly concerning STA, but it can also be considered a low energy approximation to any experimentally realistic potential. For this reason, in parallel with $V_H$, we also calculate all our results using a more experimentally-realistic sinusoidal potential \cite{TARA}.
\begin{equation}
 V_{S}(x,t)= \frac{\omega^2(t)}{2 \pi^2} \sin^2 \left(\pi \left( x-x_0(t)\right) \right) ,
\end{equation}
using the same notation as before. Note that the prefactor is chosen in such a way that $V_{H}$ is an approximation of $V_S$  around $x_0$.

\begin{figure}[tb]
\centering
\subfigure{
\centering
\includegraphics[width=0.48\linewidth]{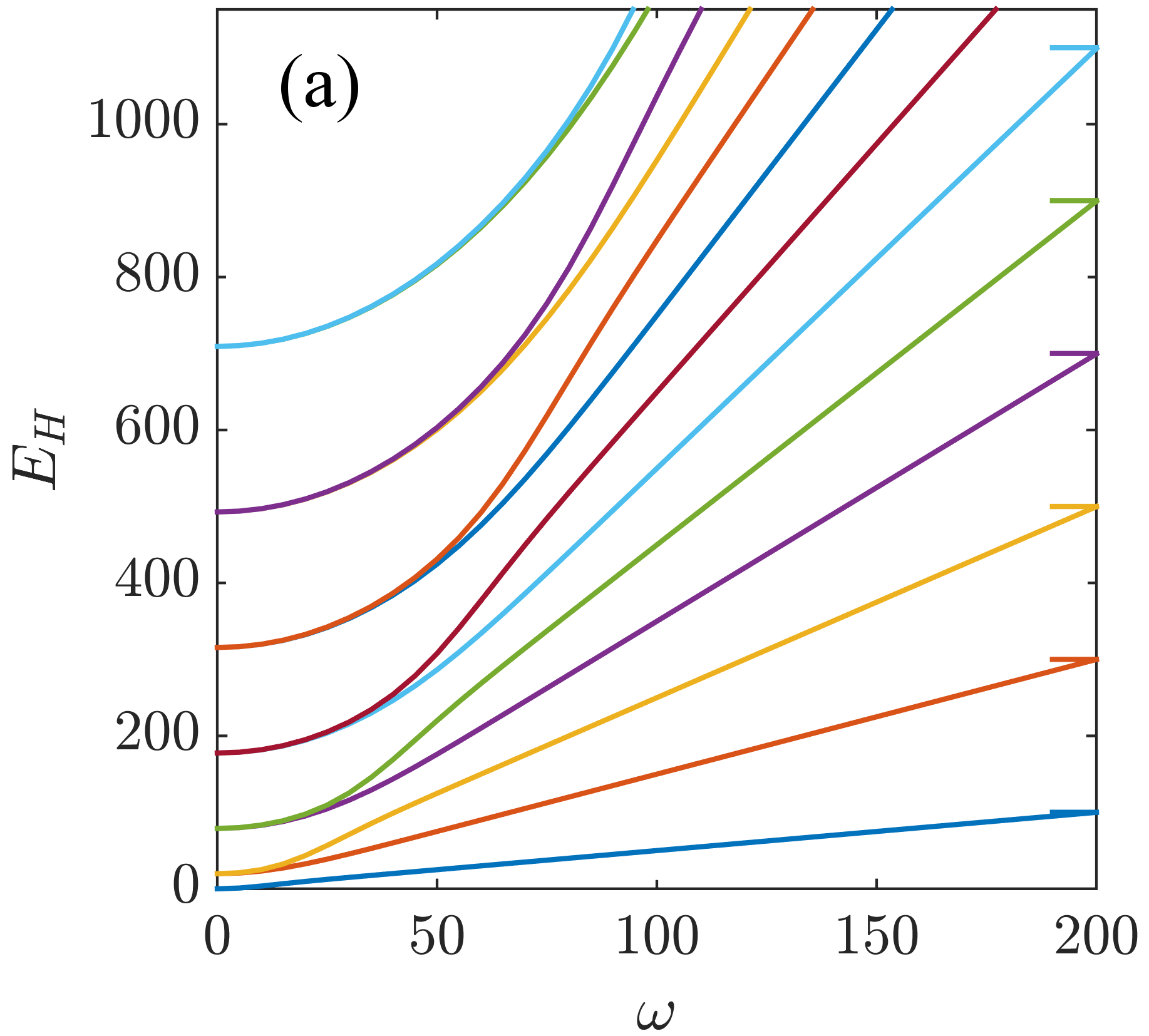}}
\subfigure{
\centering
\includegraphics[width=0.48\linewidth]{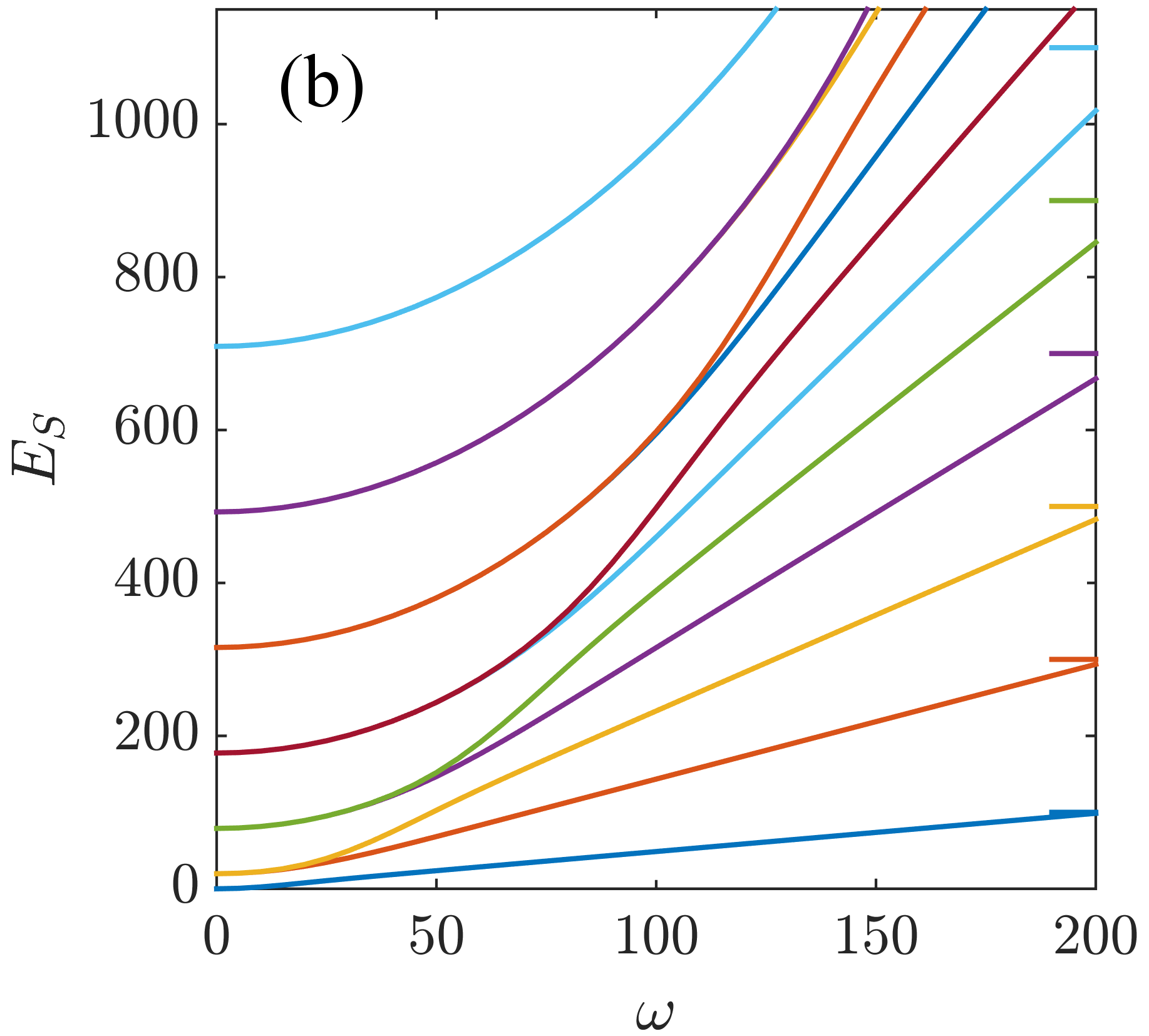}}

\caption{ Energy eigenspectrum of (a) $H_H$ and (b) $H_S$ as a function of $\omega$. 
The eigenstates continuously change from angular momentum states of energy $E_k=2\pi^2 k^2$ (with $k=0,\pm1,\ldots$) at $\omega=0$, towards harmonic-oscillator states of energy $E_n=\omega(n+1/2)$ (with $n=0,1,\ldots$) for large $\omega$.
For comparison, the horizontal lines on the vertical axis on the right give the energy levels in a harmonic potential with $\omega=200$.}
\label{fig:spectrum}
\end{figure}

To visualise the difference between the two potentials, we first compute the energy spectra of both Hamiltonians by using a straightforward discrete variable representation (DVR) method \cite{lig00,bay86} and plot the results as a function of $\omega$ in Figure \ref{fig:spectrum}.
One can see that the eigenstates at $\omega=0$ are the angular momentum states $e^{i 2 \pi k x}$, with the clockwise and counter-clockwise momentum states of opposite quantum number $k$ being degenerate. The degeneracy is lifted as $\omega$ becomes non-zero and the spectrum asymptotically approaches that of a harmonic oscillator.
Note that for the sinusoidal case, even for large $\omega$, the difference with the asymptotic harmonic spectrum increases with the quantum number $n$.

\subsection{Single particle acceleration}

\begin{figure}
\centering
\includegraphics[width=0.42\linewidth]{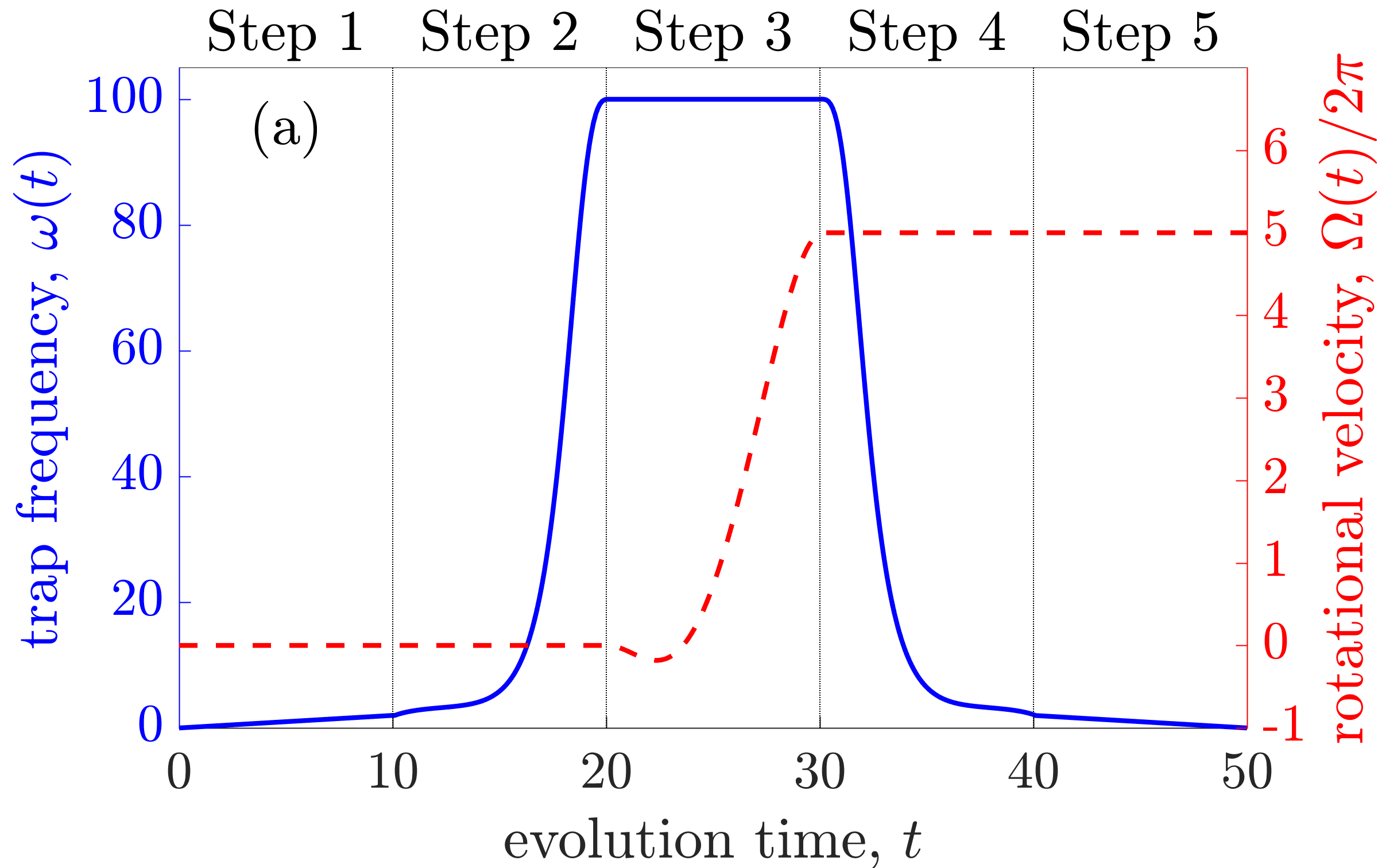} \hspace{.25in}
\includegraphics[width= 0.42\linewidth]{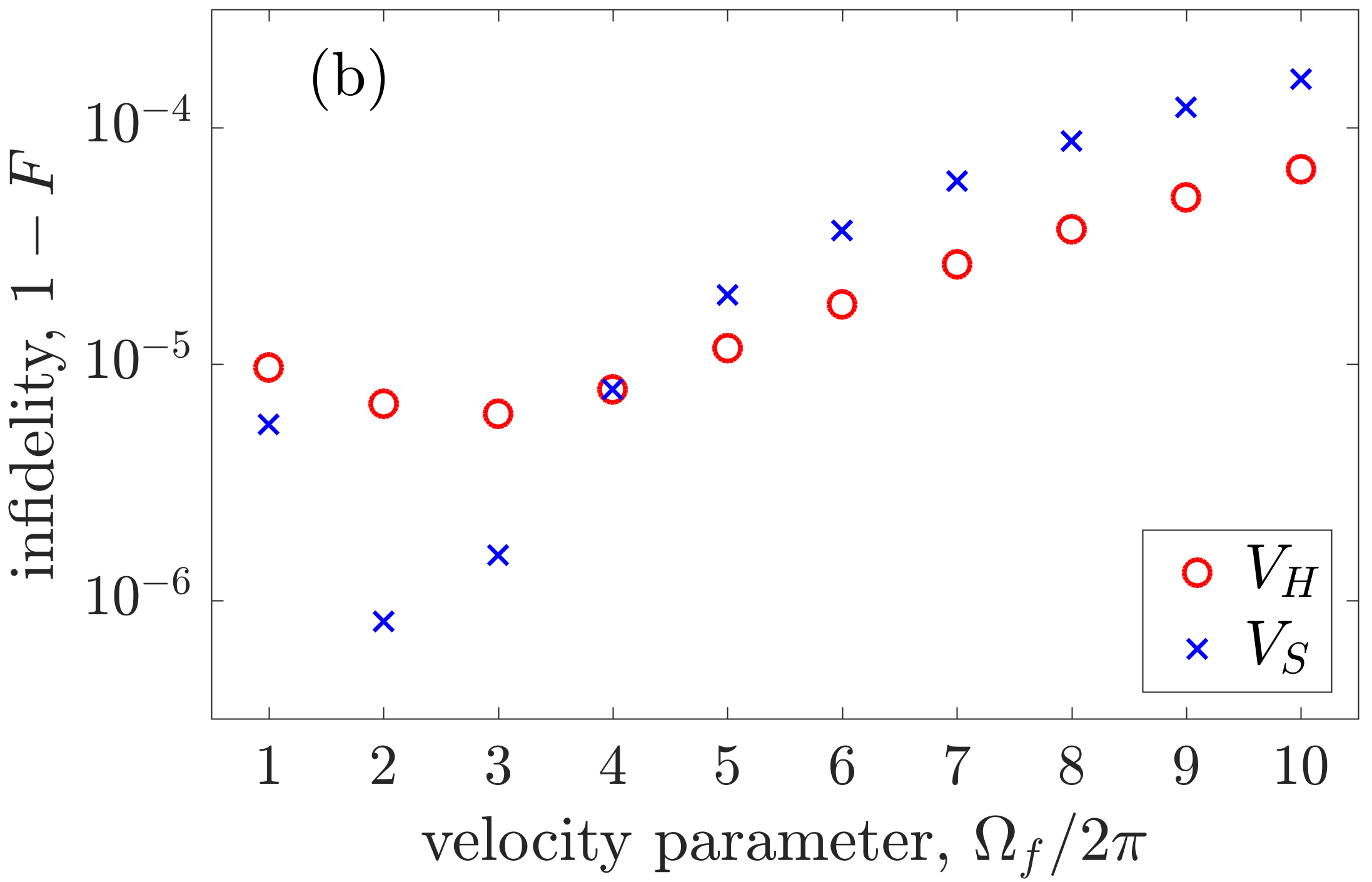}
\caption{
(a) Plot of the parameters $\omega(t)$ and the angular velocity $\Omega(t)$ for the entire protocol.
The parameters are $\omega_0=2$, $\omega_f=100$, $d=100$, each step is executed in $t_f-t_0=10$, and $\Omega_f$ is picked depending on the desired output state (here, $\Omega_f=5 \times 2\pi$).
(b) Final infidelities for $\Omega_f=1,2,\ldots,10 \times 2\pi$ for $V_H$ (dotted blue line) and $V_S$ (solid red line).
The rest of parameters are as shown in (a).
}
\label{fig:final+param}
\end{figure}

In the following, we first show results from numerically simulating our protocol for a single particle, where the free parameters of the protocol and the lengths of the different steps were chosen to allow for high fidelities.
Note that the STA raising/lowering times can in principle be made arbitrarily short for fixed $\omega_f$, unlike the adiabatic or the accelerating steps. 
In Figure~\ref{fig:final+param}(a) we show the values for $\omega(t)$ and $\Omega(t)$ which define our chosen protocol, and in Figure~\ref{fig:final+param}(b) we show the infidelities for the state preparation of plane waves $e^{i \Omega_f x}$ with $\Omega_f=1\ldots10 \times 2 \pi$.
One can see that the even for large amount of angular momentum the fidelities remain very high, for both the harmonic and the sinusoidal potential.

\subsection{Tonks--Girardeau gas acceleration}

We now consider the multi-particle case in the TG regime in order to prepare a NOON state.
As this protocol does not include a $\delta$-barrier, the target NOON state is slightly different from the one considered in the optimal control case.
It corresponds to the state originating from adiabatically removing the barrier once the system is rotating at velocity $\Omega_f=\pi$.
Similarly, the initial states for the particles will be eigenstates of free space, which are simply plane waves $e^{i 2 \pi k x}$ with integer $k$.
Since the states with $\pm k$ are degenerate, however, it is equally valid to consider the initial eigenstates
\begin{eqnarray}
&&\phi^i_0(x)=1,\\
&&\phi^i_{2l-1}(x)=\frac{1}{\sqrt 2} \left( e^{i 2 \pi l x}-e^{-i 2 \pi l x} \right)= i \sqrt{2} \sin(2 l \pi x), \\
&&\phi^i_{2l}(x)= \frac{1}{\sqrt 2}\left( e^{i 2 \pi l x}+e^{-i 2 \pi l x} )\right) = \sqrt{2} \cos(2 l \pi x),
\end{eqnarray}
for $l=1,2,\ldots$.
Those states have the property of having a total angular momentum of zero and are well suited for our STA protocol.
When an odd number of particles occupies the lower eigenstates, the $\sin$/$\cos$ pairs are guaranteed to be both populated.

For $\Omega_f=\pi$, the plane wave of quantum numbers $k+1$ and $-k$ are degenerate and we can construct the target states
\begin{eqnarray}
&&\phi^t_{2l}(x)=\frac{1}{\sqrt 2} \left( e^{i 2 \pi (l+1) x} + e^{-i 2 \pi l x} \right) = \sqrt{2} \cos[(2l+1) \pi x] e^{i \pi x} , \\
&&\phi^t_{2l+1}(x)=\frac{1}{\sqrt 2} \left( e^{i 2 \pi (l+1) x}-e^{-i 2 \pi l x} \right) = i \sqrt{2} \sin[(2l+1) \pi x]e^{i \pi x} ,
\end{eqnarray}
for $l=0,1,2,\ldots$. Those states, of total angular momentum $\pi$, are very similar to the the eigenstates considered in the previous sections and NOON states can be constructed from them. 

\begin{figure}
\centering
\subfigure{
\centering
\includegraphics[width= 0.45\linewidth]{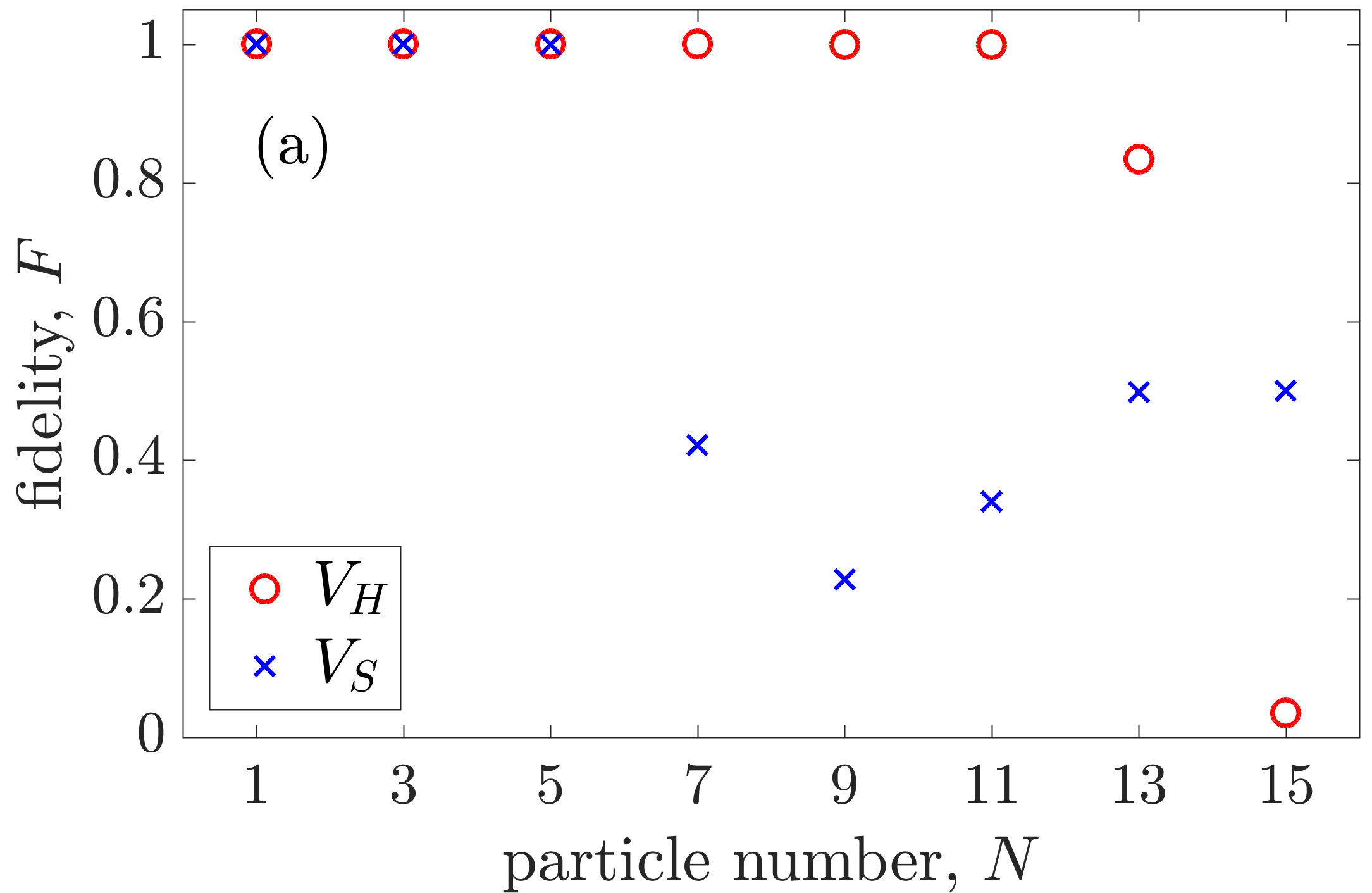} }
\subfigure{
\centering
\includegraphics[width= 0.45\linewidth]{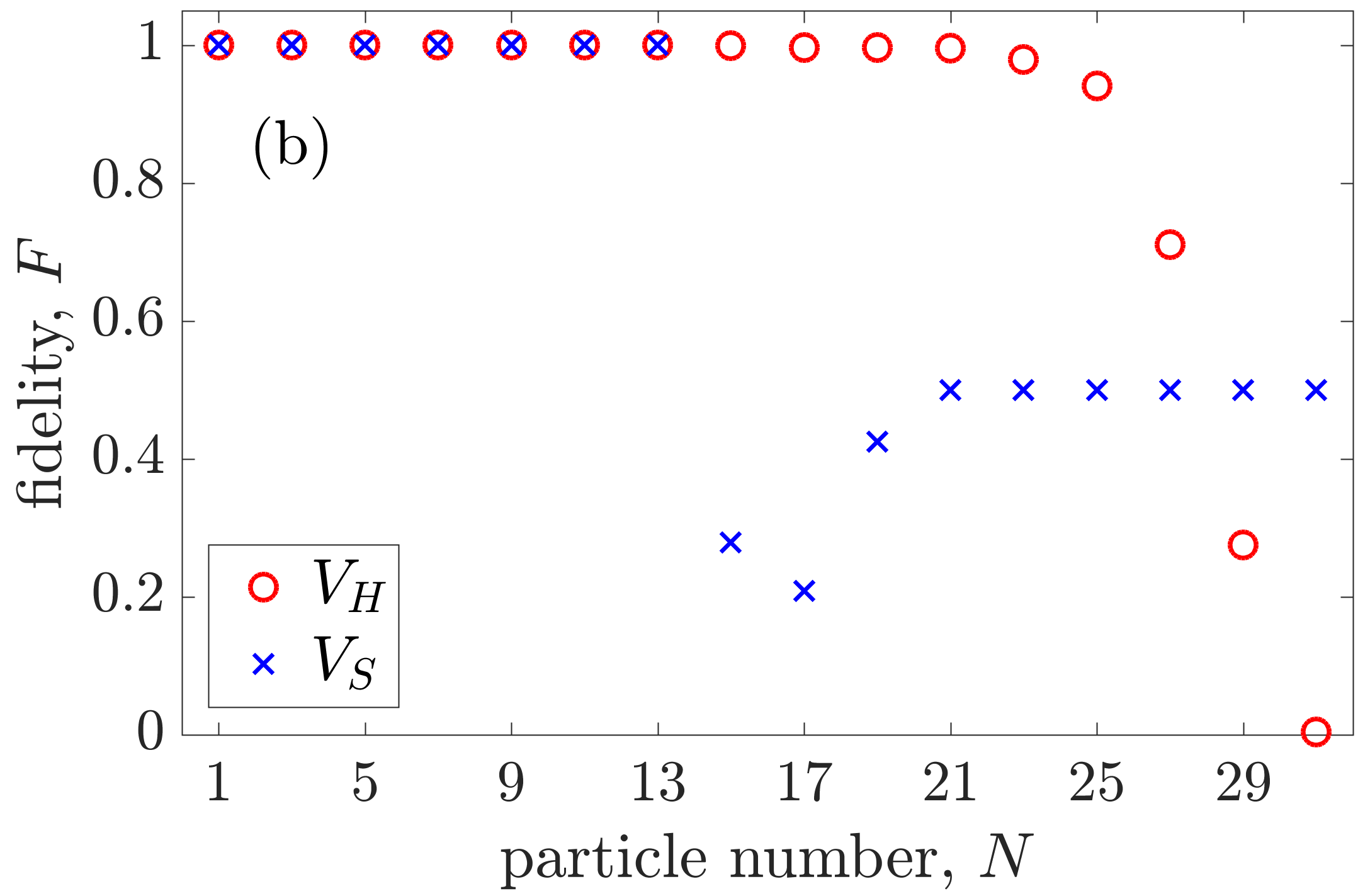}}
\caption{ Final fidelities $F$ of TG states of increasing particle number for the protocol shown in Figure \ref{fig:final+param}(a) with $\Omega_f = \pi$ for $V_H$ (red circle) and $V_S$ (blue cross). Plot (a) shows the evolution for the protocol with $\omega_f = 100$ and (b) with $\omega_f = 200$.}
\label{fig:TG-STA}
\end{figure}

An initial state $\ket{\phi^i_l}$ can be brought to the target state $\ket{\phi^t_l}$ (up to a possible shift in position) with high fidelity using our STA protocol.
This process also works for TG gases, and we show shown in Figure \ref{fig:TG-STA} the fidelity of TG gases of increasing (odd) particle numbers $N$ submitted to our protocol.  
In Figure \ref{fig:TG-STA}(a) for the harmonic potential, the fidelities remain very high (infidelities of the order of $10^{-4}$) for $N \leq 11$ after which they decrease. This is due to the finite maximum height of the harmonic potential which affects the effectiveness of the STA acceleration scheme. The fidelities can be improved by increasing the maximum trapping frequency $\omega_f$ as we demonstrate in Figure \ref{fig:TG-STA}(b) where the value of $\omega_f$ is doubled and the fidelities remain high until $N \leq 21$. 
When using the sinusoidal potential, the particle number where the fidelities drop is lower than with the  harmonic potential, although that particle number is increased when the trapping frequency $\omega_f$ is increased.

\section{Conclusion}

We have investigated the possibility of creating NOON states with ultracold quantum gases on a ring in the TG regime by using optimal control and STA techniques. With these techniques, we may evolve our system into high fidelity NOON states nonadiabatically.

In the case of optimal control, we used the CRAB technique with the Nelder--Mead minimisation method, for both a single particle and multiple particles by modifying either the potential barrier strength, its rotational velocity, or both.
By using the CRAB algorithm, we have shown that in all cases NOON states can be generated in finite time and with high fidelity.
In particular, we have shown that it is sufficient to optimise for the particle closest to the Fermi edge to achieve high TG fidelities. 
In a second approach, we have generalised two known STA techniques to a ring system, and shown that STA techniques may also be used to create rotational states with high fidelity for both single particles and strongly-correlated TG gases. The STA protocol we have applied is composed of five steps described in section 4, where only the first and final steps must be completed adiabatically. Thus we have demonstrated that it is also possible to implement STA techniques on a 1-dimensional ring system and generate NOON states between ultracold bosons without a potential barrier in the end. 

The results presented here clearly  show that it is possible to create macroscopic superposition states in TG gases on experimentally-realistic timescales. They  may therefore lead to a method of generating NOON states on a ring of ultrocold atoms for use in quantum information systems.

This work was supported by the Okinawa Institute of Science and Technology Graduate University.

\hrulefill

\end{document}